\documentclass[conference]{IEEEtran}
\pdfoutput=1
\usepackage{cite}
\usepackage{amsmath,amssymb,amsfonts}
\usepackage{algorithmic}
\usepackage{graphicx}
\usepackage{textcomp}
\usepackage{url}
\def\BibTeX{{\rm B\kern-.05em{\sc i\kern-.025em b}\kern-.08em
    T\kern-.1667em\lower.7ex\hbox{E}\kern-.125emX}}
\begin{document}
\title{\textit{In vivo} 4D x-ray dark-field lung imaging in mice}
\author{Ying Ying How$^{1^{*}}$, Nicole Reyne$^{2,3}$, Michelle K. Croughan$^{1}$, Patricia Cmielewski$^{2,3}$, Daniel Batey$^{4}$, Lucy F. Costello$^{1}$, \\ 
Ronan Smith$^{2,3}$, Jannis N. Ahlers$^{1}$, Marian Cholewa$^{5}$, Magdalena Kolodziej$^{6}$, Julia Duerr$^{7,8,9,10}$, Marcus A. Mall$^{7,8,9,10}$, \\
Marcus J. Kitchen$^{1}$, Marie-Liesse Asselin-Labat$^{4}$, David M. Paganin$^{1}$, Martin Donnelley$^{2,3}$$^{**}$, and Kaye S. Morgan$^{1}$$^{**}$
\\ \\ 
\normalsize $^{1}$ School of Physics and Astronomy, Monash University, Clayton 3800, Australia\\
\normalsize $^{2}$ Adelaide Medical School and Robinson Research Institute, University of Adelaide, North Terrace, Adelaide, Australia \\
\normalsize $^{3}$ Respiratory Medicine, Women's and Children's Hospital, North Adelaide 5006, Australia\\
\normalsize $^{4}$ Personalised Oncology Division at Walter and Eliza Hall Institute of Medical Research, Melbourne, Australia\\
\normalsize $^{5}$ Department of Biophysics, Uniwersytet Rzeszowski, Rzeszow, Poland\\
\normalsize $^{6}$ Institute of Medical Sciences, College of Medical Sciences, Uniwersytet Rzeszowski, Rzeszow, Poland\\
\normalsize $^{7}$ Department of Pediatric Respiratory Medicine, Immunology and Critical Care Medicine, Charité \\
- Universitätsmedizin Berlin, Berlin, Germany\\
\normalsize $^{8}$ German Center for Lung Research (DZL), Associated Partner Site, Berlin, Germany\\
\normalsize $^{9}$ German Center for Child and Adolescent Health (DZKJ), partner site Berlin, Berlin, Germany\\
\normalsize $^{10}$ Berlin Institute of Health at Charité‑Universitätsmedizin Berlin, Berlin, Germany\\
\normalsize  $^{*}$ ying.how1@monash.edu \\
\normalsize  $^{**}$ Equal senior authors}

\maketitle

\begin{abstract}
X-ray dark-field imaging is well-suited to visualizing the health of the lungs because the alveoli create a strong dark-field signal. However, time-resolved and tomographic (i.e.,~4D) dark-field imaging is challenging, since most x-ray dark-field techniques require multiple sample exposures, captured while scanning the position of crystals or gratings. Here, we present the first \textit{in vivo} 4D x-ray dark-field lung imaging in mice. This was achieved by synchronizing the data acquisition process of a single-exposure grid-based imaging approach with the breath cycle. The short data acquisition time per dark-field projection made this approach feasible for 4D x-ray dark-field imaging by minimizing the motion-blurring effect, the total time required and the radiation dose imposed on the sample. Images were captured from a control mouse and from mouse models of muco-obstructive disease and lung cancer, where a change in the size of the alveoli was expected. This work demonstrates that the 4D dark-field signal provides complementary information that is inaccessible from conventional attenuation-based CT images, in particular, how the size of the alveoli from different parts of the lungs changes throughout a breath cycle, with examples shown across the different models. By quantifying the dark-field signal and relating it to other physical properties of the alveoli, this technique could be used to perform functional lung imaging that allows the assessment of both global and regional lung conditions where the size or expansion of the alveoli is affected. 
\end{abstract}

\begin{IEEEkeywords}
computed tomography, \textit{in vivo} imaging, single-grid dark-field imaging, time-resolved, phase contrast
\end{IEEEkeywords}

\section{Introduction}
\label{sec:introduction}
\IEEEPARstart{D}{ark-field} imaging is a novel x-ray imaging modality that has emerged in recent decades. While the attenuation and phase shift signals originate from how much the x-ray wavefield is absorbed and phase-shifted by the sample, respectively, the dark-field signal predominantly arises from the small- or ultra-small-angle scattering of the x-rays (SAXS/USAXS) by unresolved sample microstructures. These three signals provide independent channels of complementary information about the sample. In particular, the attenuation signal is related to the imaginary part of the complex refractive index, \(\beta\) \cite{paganin2006}, whereas the phase shift signal is related to the real part of the refractive index decrement, \(\delta\) \cite{paganin2006}, and the dark-field signal is related to the sample microstructure properties such as the microstructure size, packing density and material.

The dark-field signal has been extracted using a number of imaging techniques, such as analyzer-based imaging (ABI) \cite{rigon2007_abi_3}, grating interferometry (GI) \cite{pfeiffer2008}, edge illumination (EI) \cite{endrizzi2014ei,endrizzi2017x}, speckle-based imaging \cite{berujon2012_sb,zanette2014,zdora2017x,pavlov2020x,alloo2022dark,beltran2023fast}, propagation-based imaging (PBI) \cite{gureyev2020,leatham2023x,leatham2024x,ahlers2024}, and single-grid imaging \cite{wen2010,morgan2013,croughan2022directional}. Quantitative dark-field signals have also been successfully extracted from ABI \cite{kitchen2020emphysema}, GI \cite{bech2010_coeff,lynch2011_coeff,gkoumas2016,prade2016,harti2017}, EI \cite{modregger2017} and, recently, single-grid imaging \cite{how2022quantifying,how2023quantification}. The dark-field signal can then be related to sample properties via quantities such as the kurtosis of the angular scattering distribution \cite{modregger2017}, the linear diffusion coefficient, \(\mu_{d}\) \cite{bech2010_coeff,lynch2011_coeff,gkoumas2016}, the correlation length, \(\xi_{corr}\) \cite{yashiro2010origin,prade2016,harti2017} and the dark-field scattering angle \cite{kitchen2020emphysema,how2022quantifying,how2023quantification}. 

An x-ray dark-field signal can be useful in various fields, including biomedical imaging, materials science and security screening. Safety or industrial applications include detecting and/or identifying goods that come in powder forms, such as drugs or explosives \cite{miller2013phase,partridge2022enhanced}, and imaging industrial parts made from carbon fibres \cite{valsecchi2020}. Some possible biomedical imaging applications include imaging kidney stones of different compositions and microscopic morphology for classification \cite{scherer2015non, niemann2021classification}, imaging breast tissues with microcalcifications for early detection of cancer \cite{michel2013dark, wang2014non, forte2020can, aminzadeh2022}, and imaging bones to diagnose osteoporosis or undisplaced fractures \cite{gassert2023dark,schaff2024feasibility}. Several animal studies have also shown that lung diseases that lead to a change in size or structure of the alveoli \cite{schleede2012emphysema,yaroshenko2013pulmonary,kitchen2020emphysema,scherer2017x, hellbach2015vivo,hellbach2017x} can result in a change in the dark-field signal from the lungs. 

Dark-field imaging of the lungs has attracted particular attention because the lungs generate a stronger dark-field signal than any other part of the human body. The lungs are made up of hundreds of millions of small air sacs called alveoli, which is where gas exchange occurs. When x-rays pass through these alveoli, which are not typically resolved in an x-ray image, they experience multiple refraction and/or scattering at the interfaces of the alveoli tissue and air. This scattering of x-rays then gives rise to the dark-field signal, and depending on the size and the number of alveoli that the x-rays pass through, the strength of the dark-field signal can differ \cite{schleede2012emphysema}. 

Dark-field lung imaging using grating interferometry has moved particularly quickly towards clinical application and has shown promising results. Recently, the diagnostic capability of dark-field radiographs has been demonstrated on healthy individuals \cite{gassert2021x}, COVID-19-pneumonia patients \cite{frank2022dark}, and chronic obstructive pulmonary disease (COPD) patients with emphysema \cite{willer2021x}. Dark-field lung radiography has also been shown to perform better than conventional lung radiography for diagnosing and staging emphysema \cite{urban2023dark}. A human-sized prototype grating interferometry Computed Tomography (CT) system has also been built \cite{viermetz2022dark} in preparation for performing three-dimensional dark-field lung imaging in a clinical setting.

In the case of \textit{in vivo} small-animal studies, most dark-field lung imaging has been performed either with images acquired over several breaths \cite{meinel2013diagnosing,yaroshenko2013pulmonary,hellbach2015vivo}, which results in a motion-blurring effect, or have required the sample to remain in a breath-hold position \cite{gromann2017vivo}. To resolve changes through the breath, dynamic \textit{in vivo} dark-field imaging on mice has been performed with a grating interferometer \cite{gradl2018dynamic}. In that study, the mouse was continuously ventilated while the analyser grating was stepped over 7 positions, so that images taken at different grating positions (but the same breath point) could be used to retrieve the dark-field signal projections at each breath point. 

However, a two-dimensional projection of a three-dimensional sample can obscure details and confound projected thickness with attenuation/dark-field coefficients. This can be overcome by performing CT using projections collected at multiple angles to obtain the spatial information missing from a projection. In particular, a three-dimensional dark-field image could be useful for diagnosing lung diseases by revealing the internal structures of the lungs and providing complementary information that is not accessible from conventional CT images. Furthermore, by synchronizing the image acquisition with the breathing/ventilation cycle, a four-dimensional dark-field image of the lungs can be obtained, which then allows changes in alveolar dimensions to be tracked throughout a breath cycle, mapped across each part of the lungs. This is a type of functional imaging of the lungs, which can allow the condition of the lungs to be assessed more accurately than conventional methods such as spirometry, which average the lung health across the whole lung into one measure. 

To perform \textit{dynamic} three-dimensional (i.e.,~four-dimensional, with time being the fourth dimension) dark-field imaging, it is helpful to have a technique that can extract the dark-field signal from a single sample exposure, to minimize the data acquisition time. One potential candidate for this is the single-grid imaging technique \cite{wen2010,morgan2011_grid}, which is similar to grating interferometry but uses only one grid, the image of which is resolved at the detector, avoiding the need to step the grid. Three complementary signals can be extracted by comparing the sample-and-grid image to the grid-only image. The single-grid imaging technique utilizes a relatively simple setup (see Fig.~\ref{mice_setup}) compared to other dark-field imaging techniques, and it does not require any calibration or alignment prior to the data acquisition process. The short data acquisition time associated with this technique can also reduce any potential motion-blur in the image and reduce the radiation dose delivered to the sample, which is one of the key considerations in clinical imaging. The sensitivity of the imaging system to the dark-field signal can also be tuned by simply changing the sample-to-detector propagation distance and the pixel size, where a higher sensitivity can be achieved with a smaller pixel size or a larger propagation distance. 

Recently, How \textit{et al.} developed an algorithm to extract and quantify the dark-field signal from single-grid imaging \cite{how2022quantifying} and related the dark-field signal to the sample microstructure size \cite{how2023quantification}. This was achieved by modelling the grid intensity pattern, with and without the presence of the sample, as two sinusoids and performing curve-fitting on the cross-correlation results between the two sinusoids. The algorithm has also been successfully applied to images taken using speckle-based imaging, which is similar to single-grid imaging, where the grid is simply replaced with a piece of sandpaper \cite{morgan2013, berujon2012_sb}, with the results shown in Section 6.2 of How and Morgan~\cite{how2022quantifying}.  

Here we present the first dynamic \textit{in vivo} dark-field lung CT imaging of mice. This was achieved using single-grid imaging and synchronizing the image acquisition with the breath cycle of the mice. In total, 8 sets of $180^{\circ}$ sample-and-grid projection images were obtained from a single scan, where each set corresponds to a different equally spaced timepoint within the breath cycle. The projection images were analyzed using the algorithm developed in \cite{how2022quantifying} to extract the attenuation, phase shift, and dark-field signals. CT reconstructions were then performed on the extracted images to obtain a three-dimensional lung volume of respective signals at each timepoint. Here, a \(\beta\)-ENaC-Tg mouse \cite{mall2004increased,mall2008development,zhou2017airway,wielputz2011vivo,mccarron2021animal,zhu2022muct,reyne2024} with muco-obstructive disease and emphysema, a mouse with induced lung cancer, and a wild-type mouse were imaged. With the four-dimensional dark-field data, we were able to measure and track how the strength of the dark-field signal from different voxels across the lung changes throughout a breath cycle across these three different mouse models. By quantifying the dark-field signals and relating them to the properties of the alveoli, this technique has the potential to provide useful diagnostic information for lung diseases.

\section{Experimental Setup}
\subsection{Imaging system}
\begin{figure}[!t]
\centerline{\includegraphics[width=\columnwidth]{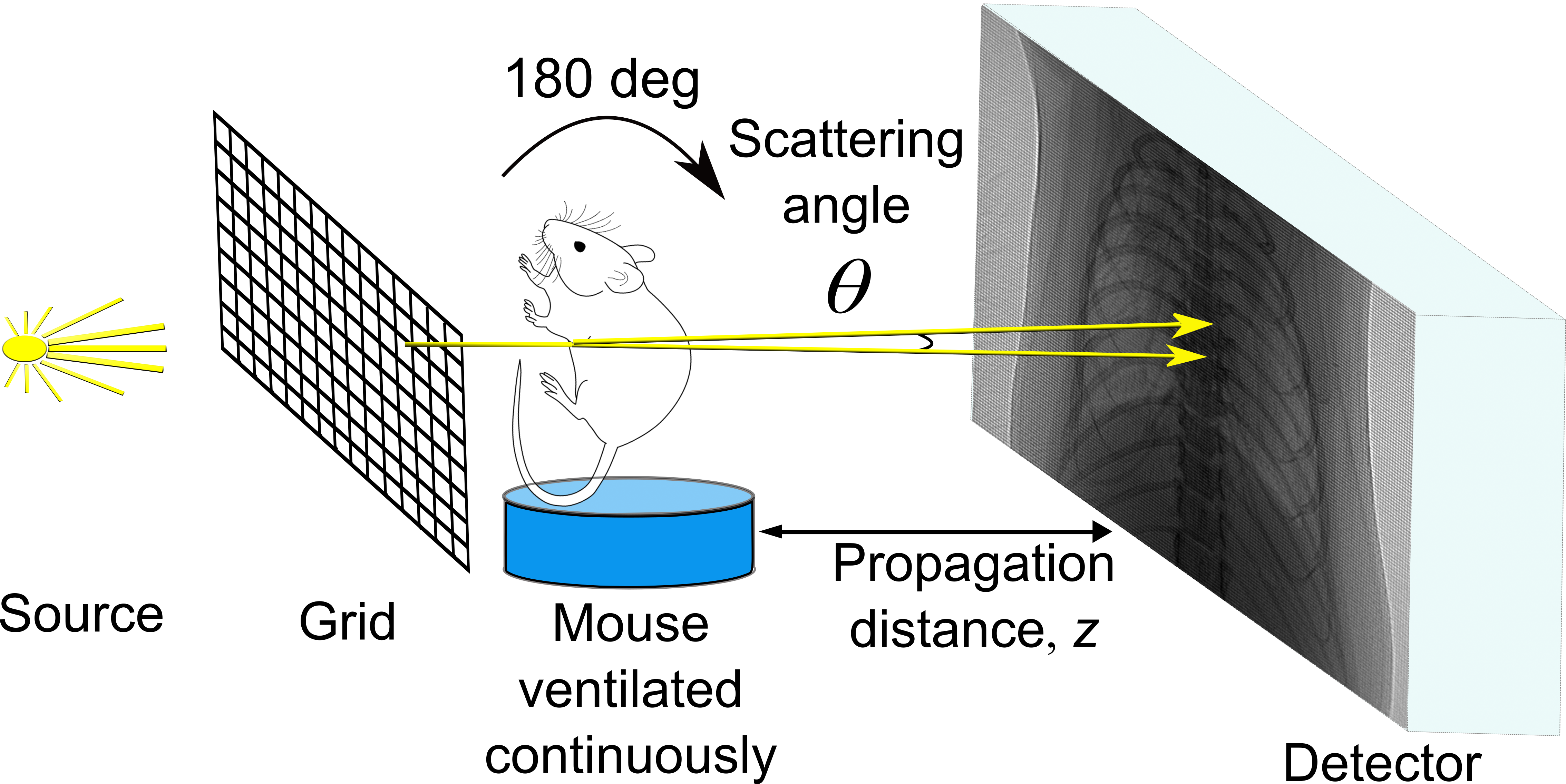}}
\caption{The experimental single-grid imaging setup for \textit{in vivo} 4D dark-field lung imaging of mice, used here with a propagation distance of \(z\)=3.5 m. The mouse was mechanically ventilated while being rotated $180^\circ$ to perform a CT. The image acquisition procedure was synchronized with the ventilation cycle.}
\label{mice_setup}
\end{figure}

The experiment setup is shown in Fig.~\ref{mice_setup}, and consists of a typical single-grid imaging setup at a synchrotron. The experiment was performed at the Australian Synchrotron Imaging and Medical Beamline (IMBL), where the energy of x-rays was set at 25~keV. An attenuating grid (a geological stainless steel sieve with square holes of width 90~{\textmu}m and wires 48~{\textmu}m thick, resulting in an effective period of 138~{\textmu}m) was placed 50~cm upstream of the sample (i.e.,~as close as possible). A pco.edge 5.5 sCMOS detector coupled to a 25~{\textmu}m thick Gadox phosphor was placed 3.5~m downstream from the sample, providing an effective pixel size of 9.7~{\textmu}m. This sample-to-detector propagation distance was decided experimentally by visually inspecting the sample-and-grid image and running trial analysis to ensure that the grid intensity pattern had high visibility and that the dark-field blurring of that pattern by the lungs was detectable (see Section \ref{optimal_setup} for details). The sample was placed on a CT stage and the attenuating grid was placed on a separate stage, enabling both to be moved in the horizontal and vertical directions remotely so that they could be moved out of the field of view to take flat-field or grid-only images. 

\subsection{Ethics Statement} 
All animal procedures were approved by the University of Adelaide (M-2023-070), Walter and Eliza Hall Institute (2023.006), and ANSTO (AS2023\_04) animal ethics committees, and were conducted in accordance with the National Health and Medical Research Council Australian Code of Practice of the Care and Use of Animals for Scientific Purposes. 

\subsection{Mouse preparation}
This proof-of-principle study focused on two lung disease models, the \(\beta\)-ENaC-Tg mouse \cite{mall2004increased,mall2008development,zhou2017airway,wielputz2011vivo,mccarron2021animal,zhu2022muct,reyne2024}, which is a model of muco-obstructive disease and emphysema, and the KRAS;p53 mutant (KP) \cite{weeden2023early} mouse model of lung cancer. The \(\beta\)-ENaC-Tg mouse (5-6 months old, at which time the emphysema phenotype is dominant) was supplied from the breeding colony at the South Australian Health and Medical Research Institute Preclinical, Imaging and Research Laboratories (SAHMRI PIRL) facility in Adelaide, Australia. The lung cancer mouse was supplied by the Walter and Eliza Hall Institute in Melbourne, Australia, and was prepared by intravenously injecting 2.5x10\textsuperscript{5} \textit{Kras\textsuperscript{G12D};p53\textsuperscript{\(\Delta\)/\(\Delta\)}} lung cancer cells (a kind gift from Dr K. Sutherland) into C57B1/6 mice 4-6 weeks prior to the experiment, as previously described \cite{weeden2023early}. The control mouse was the normal littermate from the \(\beta\)-ENaC-Tg colony. 

Mice were anaesthetized with an intraperitoneal (i.p.) injection of a mixture of 75~mg/kg of ketamine (Ceva, Australia) and 1~mg/kg medetomidine (Ilium, Australia). Once anaesthetized, the mice were prepared by performing a tracheostomy followed by cannulation with an endotracheal tube (ET; 20 Ga BD Insyte plastic cannula). The mice were then placed in an animal holder in a supine (head-high) position with both arms raised and attached to the holder to ensure a clear field of view of the chest. The fur in the chest region of the mice was not removed since the dark-field signal generated by the fur was significantly weaker than the signal generated by the lungs. A heat lamp was used throughout the imaging to ensure constant body temperature. Mice were connected to an Accuvent 200 small animal ventilator (4DMedical, Australia) configured to a peak inspiratory pressure of 14~cmH\textsubscript{2}O, positive end-expiratory pressure of 2~cmH\textsubscript{2}O, and ventilated at 120 breaths/min (170~ms inspiration and 330~ms expiration).

\subsection{Time-resolved image acquisition}\label{image_acquisition}
\begin{figure*}[!t]
\centerline{\includegraphics[width=\textwidth]{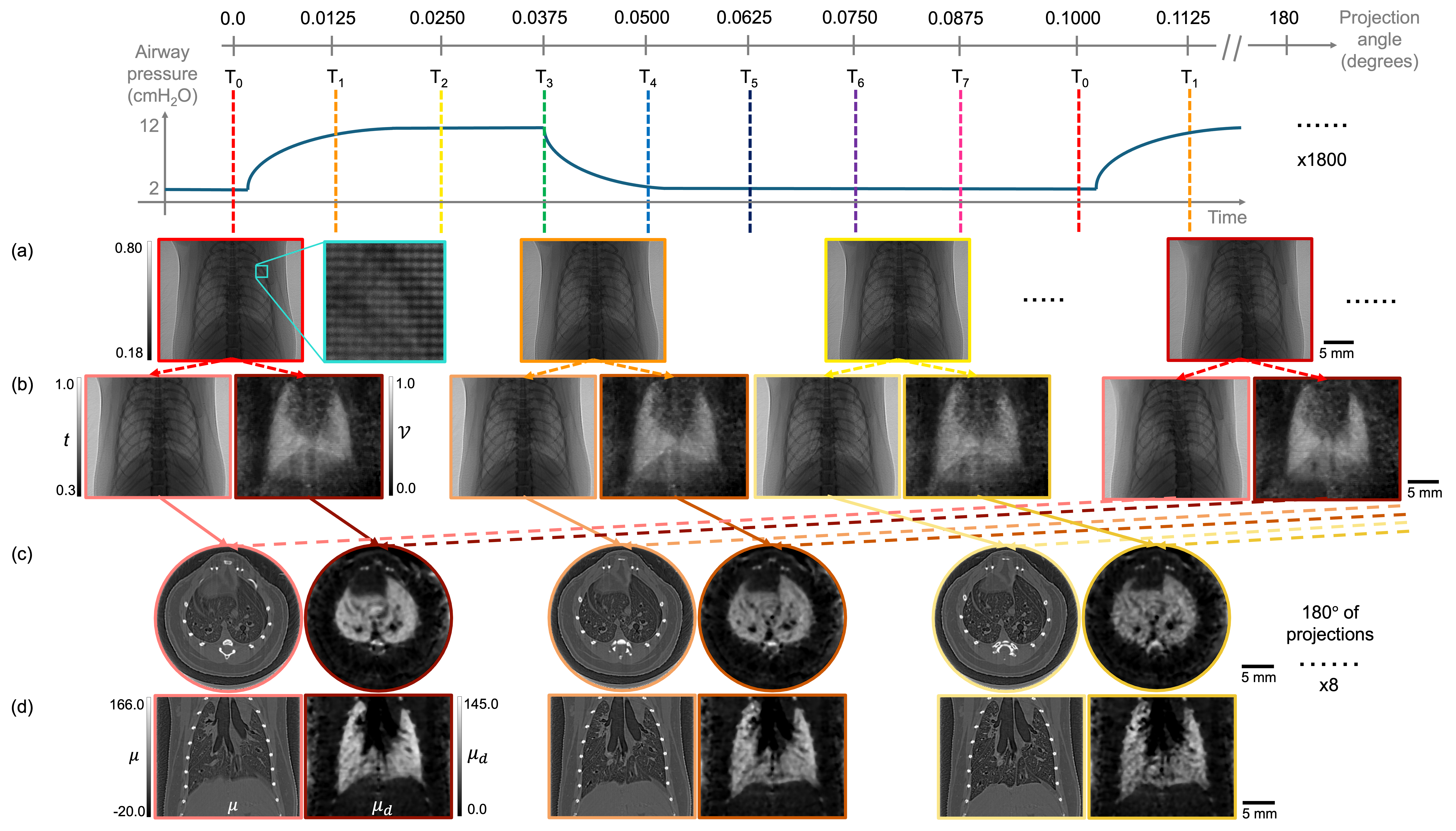}}
\caption{A schematic diagram showing the synchronized image acquisition procedure. (a) Images were taken at 8 equally-spaced points (labelled with different colors) in the breath cycle. The sample was continually rotated, allowing images to be captured across the breath cycle and across 180 degrees of projections. (b) Each projection allowed an attenuation and dark-field image to be retrieved. (c) \& (d) Selecting projections from the same breathpoint allows the reconstruction of attenuation and dark-field CTs. Full details provided in Section \ref{image_acquisition}. Note that for illustration purposes, to demonstrate a rotation of the sample, the T\textsubscript{0} image on the far right here is taken at the 100\textsuperscript{th} breath, not the second breath. All the reconstructed attenuation and dark-field CT slices in this manuscript are shown with the same greyscale. See Supplementary Video 1 for a 3D view of the images.} 
\label{breath_cycle}
\end{figure*}

Image acquisition was performed in synchronization with the ventilation cycle (see Fig.~\ref{breath_cycle}) to capture the dynamics of the lung. This approach utilizes the repetitive motion of the breathing of the mice and assumes that the lungs return to the same position at the same timepoint in each breath cycle to capture the next projection within the CT. A custom-designed timing hub \cite{morgan2020methods} used ventilator triggers to acquire images at 8 different evenly-spaced breath points, with the mouse rotated continuously. Over each breath cycle, the mouse was rotated by $0.1^\circ$, resulting in a rotation of $0.0125^\circ$ between exposures. By repeating this acquisition for 1800 breaths, 8 sets of $180^\circ$ of projections were collected, with an increment of $0.1^\circ$ between each projection image in a set, and where each of the 8 sets corresponds to a different breath point. The acquisition time for each projection was 62~ms, which consisted of 55~ms of exposure time and 7~ms of image read-out time. The total time required to perform one 4D CT scan was 15 minutes. 

A CT was taken for each mouse with the grid present. For each CT scan, the mouse was also moved out of the field of view to collect the grid-only and flat-field images. Dark-current images were taken at the end of each CT scan. Note that the flat-field images were taken with a 40~ms exposure to match the intensity levels seen in projections, and then scaled accordingly.

After imaging, the mice were humanely killed by i.p. overdose of lethabarb (150~mg/kg, Virbac, Australia). The mice remained in the imaging position after being euthanized, and the ventilator was set to maintain the lung at a constant pressure of 10~cmH\textsubscript{2}O. The sample-to-detector propagation distance was then reduced to 0.7~m and a second CT scan without the grid was performed to obtain high-resolution phase-retrieved CT slices for comparison to the lung tissue histology and the dark-field CT. Note that the lack of lung motion enabled a higher spatial resolution than possible in the dynamic dark-field CT scan.

\subsection{Histology}
After x-ray imaging was completed, the lungs were inflation-fixed \textit{in situ} using a 10\% formalin buffer at 18~cmH\textsubscript{2}O, removed and placed into formalin for at least 24 hours, and then transferred in 70\% ethanol. Samples were embedded in paraffin wax and 5~{\textmu}m slices were mounted onto slides. \(\beta\)-ENaC-Tg tissues were stained with Alcian Blue/Periodic Acid Schiff for detection of mucus, and the lung cancer mice with haematoxylin and eosin. Images were captured using an Olympus VS200 slide scanner, and visualized using QuPath software.  

\section{Analysis}
\subsection{Extracting the dark-field signal from projections}
The projection images were first sorted into their corresponding breath points and flat- and dark-current-corrected. These images (shown in Fig.~\ref{breath_cycle} (a)) were then analysed using the single-grid dark-field retrieval algorithm as described in \cite{how2022quantifying}, with a cross-correlation window size of 16 pixels to match the grid period. To minimize the computing time, the analysis was performed for every 10\textsuperscript{th} pixel in the image. Although the attenuation and dark-field images extracted (shown in Fig.~\ref{breath_cycle} (b)) in this way had a lower spatial resolution compared to the raw images, the signals extracted across the neighbouring pixels within the same period of the grid (i.e.,~16 pixels) were expected to be similar. The dark-field images were then smoothed by a median square kernel of size 5 pixels before the CT reconstruction to reduce the noise level and minimize any artefacts originating from particularly bright/dark pixels.    

\subsection{CT reconstruction}
CT scans were reconstructed using the ASTRA Toolbox \cite{van2015astra,van2016fast} implementation of filtered-back-projection with the Ram-Lak filter. The dark-field signal retrieved using the algorithm described in \cite{how2022quantifying}, \(\mathcal{V}\) (referred to as \(DF\) in \cite{how2022quantifying}), is defined as the relative change in the visibility between the sample-and-grid intensity image and grid-only intensity image. This relative change in visibility, \(\mathcal{V}\), which is the most common way to measure the dark-field signal in grating-based techniques such as grating interferometry, follows an exponential relationship with sample thickness, \(T\) (similar to the Beer-Lambert Law) \cite{bech2010_coeff,lynch2011_coeff}, 
\begin{equation}
    \mathcal{V} = \exp (-\mu_{d}T), \label{DF_CT_eqn}
\end{equation}
where \(\mu_{d}\) represents the linear diffusion coefficient/dark-field extinction coefficient. Given \eqref{DF_CT_eqn}, we reconstructed the dark-field CT of the lungs in the same way as we did for the attenuation images, shown in Fig.~\ref{breath_cycle} (c) \& (d). After the CT reconstruction, a three-dimensional median filter with a kernel size of 3 pixels was applied to the three-dimensional dark-field lung volume to minimize noise and any ring artefacts observed in the CT slices.        

\subsection{Region of Interest (ROI) measurements}  
The spatial and temporal variations in the dark-field signal could be extracted from the reconstructed three-dimensional dark-field lung volume at 8 different breath points. For example, the relative strength of the dark-field signal in the left and right lung could be measured from a given CT slice. The strength of the dark-field signal within certain voxels, or within the left, the right, or even the whole lung, could also be tracked throughout the breath cycle. Median values, rather than mean, were measured from these regions of interest, given that part of the lungs is made up of blood vessels or fully resolved airways, which do not generate dark-field signals.  

To extract time-resolved measurements at a given axial slice, the motion of the lung had to be considered. That is, because the lungs were primarily expanding in the vertical direction towards the diaphragm during inhalation, a particular feature within the lung tissue moves downwards through the axial CT slices during the breath. This resulted in some lung tissues moving in and out of the specific slice that we were interested in during the breath cycle (see Supplementary Video 2). Therefore, manual alignment of the lung tissue was performed using features within the lungs such as the blood vessels or major airways seen in the attenuation CT, in order to observe the same lung tissue throughout the breath cycle. The same alignment was then applied to the dark-field CT slices, and as a result, a given measurement was performed over CT slices with different slice numbers for each breath point, rather than with a fixed slice number. For example, the 100\textsuperscript{th} axial slice at T\textsubscript{0} matched better with the 103\textsuperscript{rd} slice at T\textsubscript{1}, the 105\textsuperscript{th} slice at T\textsubscript{2} and the 99\textsuperscript{th} slice at the last breath point, T\textsubscript{7}.  

Masks were manually drawn for the left and right lungs, enabling measurements for either side or the whole lung. The masks were traced on the attenuation CT slice at the last breath point (i.e.,~the end of exhalation) when the lung volume was the smallest. This ensured that the mask contained only lung tissue at every breath point. For more local measurements, ROIs of size 10 pixels \(\times\) 10 pixels were chosen carefully to only contain lung tissue through the breath, and neither blood vessels nor fully-resolved airways.     

\section{Results}
\subsection{Spatial variations in dark-field}
\begin{figure}[!t]
\centerline{\includegraphics[width=\columnwidth]{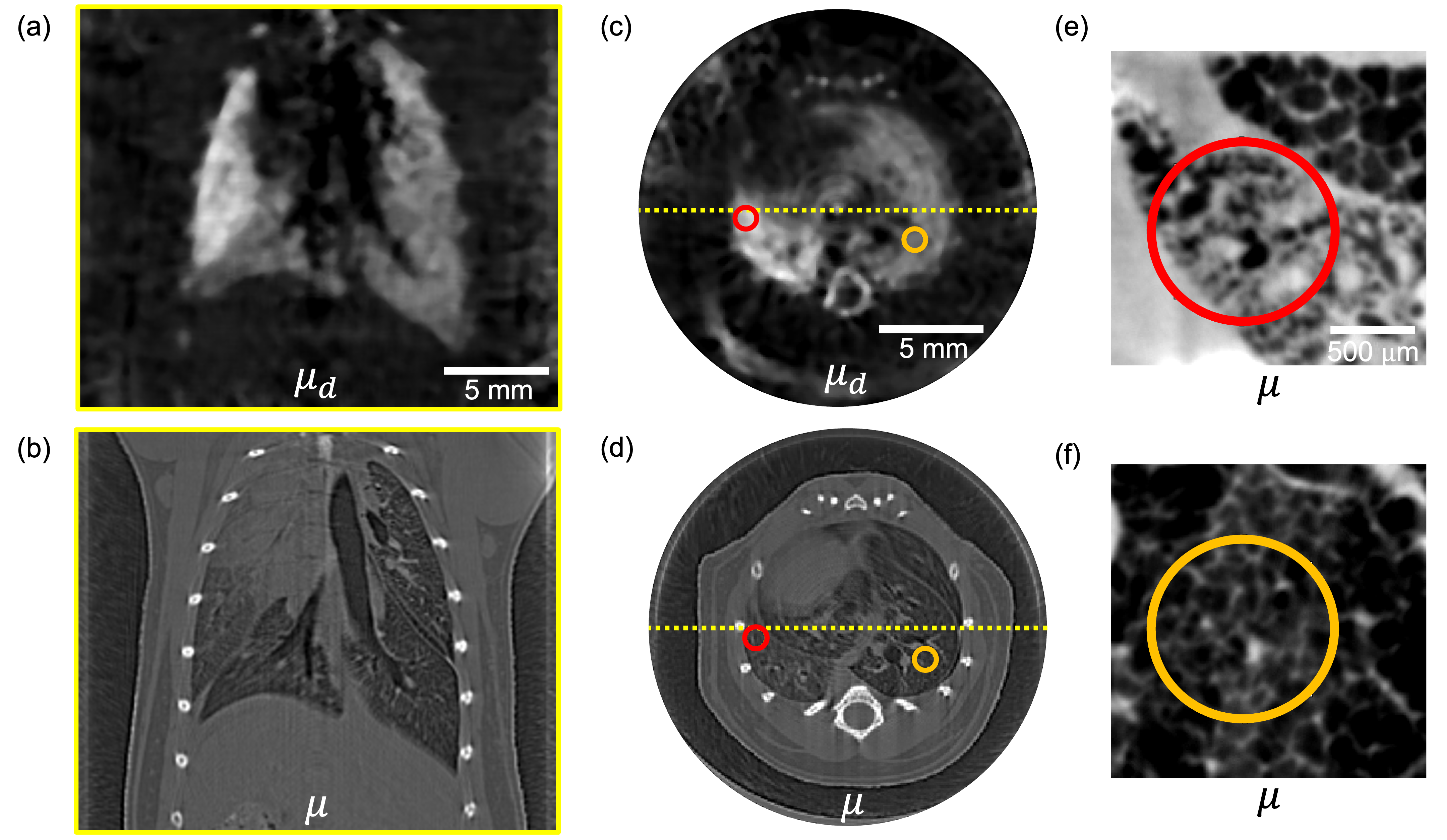}}
\caption{Representative example of the reconstructed (a) \& (c) dark-field, \(\mu_{d}\), and (b) \& (d) attenuation, \(\mu\), CT slices in the coronal and axial planes of the \(\beta\)-ENaC-Tg mouse at T\textsubscript{0}, together with (e) \& (f) the magnified images of the high-resolution phase-retrieved CT slices captured post-mortem. ROIs were selected from two different lobes of the lungs that generated visibly different dark-field signal strength, which is explained by the different lung structures seen in (e) \& (f). See Supplementary Video 3 for a time-resolved version of this figure and see Supplementary Figure S2 for the histology slices.}
\label{spatial_DF} 
\end{figure}

Fig.~\ref{spatial_DF} (a) - (d) shows the reconstructed attenuation and dark-field CT of the \(\beta\)-ENaC-Tg mouse, using signals retrieved from the single-grid dark-field retrieval algorithm, sliced at the coronal and axial planes. A clear difference in the strength of the signal between lobes was observed in the dark-field images, which was not as visible in the attenuation images. This is reflected in an increase of the contrast-to-noise ratio (CNR) between the two lobes from 2.4 in the retrieved attenuation CT slice (Fig.~\ref{spatial_DF} (d)) to 5.5 in the retrieved dark-field CT slice (Fig.~\ref{spatial_DF} (c)). This difference in the dark-field signal suggests that the size of the alveoli in the lobes was different, where smaller and more tightly-packed alveoli generate a stronger dark-field tomographic signal than large and/or loosely-packed alveoli. From the magnified image of the selected regions of interest in post-mortem high-resolution CTs (Fig.~\ref{spatial_DF} (e) \& (f)), it was observed that the alveoli in the right lobe of the lung (labelled in orange) are indeed larger compared to the left lobe (labelled in red), which is consistent with the dark-field signal. 

\subsection{Temporal variations in dark-field}
\begin{figure}[!t]
\centerline{\includegraphics[width=\columnwidth]{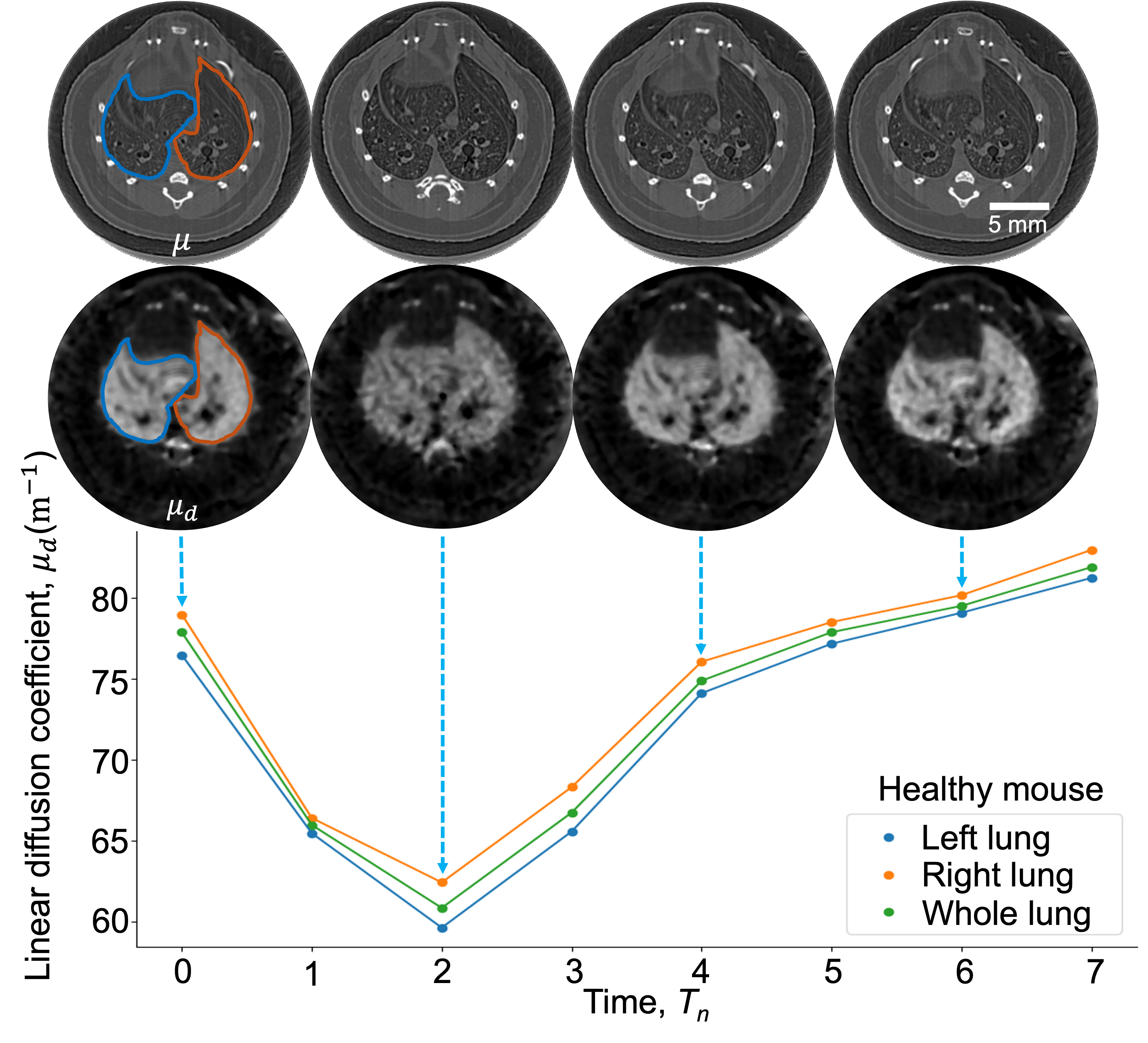}}
\caption{The change in the dark-field signal from the lungs of a control mouse throughout the breath cycle. The dark-field signal was measured from a slice of thickness 97~\textmu{m} taken from the middle part of the lungs. The measurement for the whole lung (green) was taken from both the regions highlighted in blue (left lung) and orange (right lung). At T\textsubscript{2}, when the lungs are most inflated, the dark-field image has a more porous appearance, suggesting that airways that were unresolved at T\textsubscript{0} are now inflated and resolved (e.g.,~top right of the axial slice).}
\label{temporal_DF}
\end{figure}

Fig.~\ref{temporal_DF} shows how the dark-field signal from the lungs of a normal control mouse changes through a breath cycle. The median dark-field signal measured from the left (blue), right (orange) and whole (green, i.e.,~combining left and right) lungs follow a similar trend with time. From the beginning of the breath cycle, the strength of the dark-field signal decreased and reached a minimum value at timepoint T\textsubscript{2}, which corresponds to the period of inhalation (as shown in Fig.~\ref{breath_cycle}). During exhalation, the strength of the dark-field signal gradually returned to its initial level. The change in the dark-field signal suggests that the alveoli were expanding during the first part of the breath cycle, and after reaching maximum expansion (see the dark-field CT image at T\textsubscript{2} in Fig.~\ref{temporal_DF} and in Fig.~\ref{breath_cycle}), the alveoli were relaxing/contracting for the remainder of the breath cycle. It is worth noting that while the dark-field images at different breath points were substantially different from each other, the attenuation images were relatively similar to each other. This suggests that the dark-field image can complement the attenuation image, which only shows the fully resolved features, by revealing information about unresolved sample features like the alveoli.  

\subsection{Spatial and temporal variations in dark-field}

\begin{figure}[!t]
\centerline{\includegraphics[width=\columnwidth]{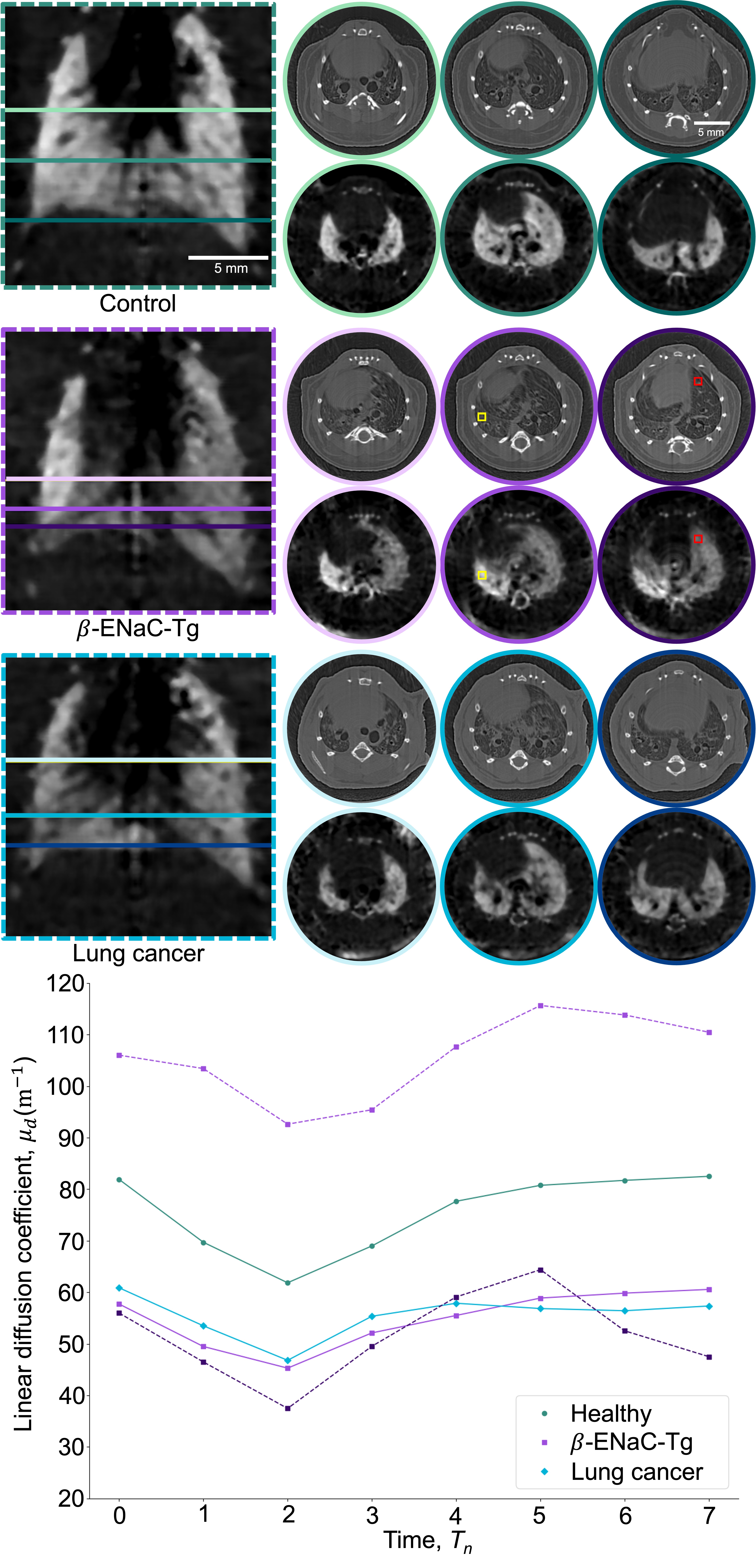}}
\caption{Dark-field signals from different parts of the lungs at different breath points across different models. Top: Attenuation and dark-field axial CT slices retrieved at T\textsubscript{0} from different parts of the lungs, as indicated by the corresponding colored horizontal lines in the dark-field coronal slices. Bottom: Plot showing how the dark-field signal of the whole lung in the middle slice (solid lines) or specific voxels of interest (purple dotted line for yellow box and dark-purple dotted line for red box) changes throughout the breath cycle. See Supplementary Video 4 for a time-resolved version of this figure and Supplementary Videos 5-7 for the CTs of the breathing control, \(\beta\)-ENaC-Tg, and cancer-affected mice.}
\label{spatial_temporal_DF}
\end{figure}

With this 4D technique, we were able to extract time-resolved measurements in highly-localised regions, as shown in Fig.~\ref{spatial_temporal_DF}. The dark-field CT images show variations in dark-field signal across the lungs, suggesting some variations in the size or the number of alveoli. The dark-field signals from different disease models also follow different trends. In particular, the \(\beta\)-ENaC-Tg and cancer lungs (purple and blue) generate weaker dark-field signals and smaller relative changes in magnitude compared to the control lung (green). 

The observed decrease in the strength of the dark-field signal in the \(\beta\)-ENaC-Tg lung corresponds to the increase in the alveoli size (as shown in Figure S2 of the Supplementary Information). In the cancerous lung, the observed decrease in dark-field signal may be attributed to tumour lesions obstructing airflow and hence affecting the aeration of the alveoli, and/or a decrease in the fraction of the lung tissue made up of alveoli (as shown in Figure S3 of the Supplementary Information).

The variations in dark-field signal across the breath cycle are also reduced for the disease models, suggesting reduced elasticity of the alveoli and/or obstructed airflow. By focusing on specific regions of interest, we can obtain localised lung health measurements that are lost in an averaged measurement across the whole lung or potentially even in projection. This is shown by the stronger dark-field signal measured from the yellow ROI in the left lobe of the \(\beta\)-ENaC-Tg lungs, which suggests the presence of more densely-packed alveolar tissue, particularly when compared to the red ROI in the right lobe. These results show the potential of time-resolved dark-field CT for providing more accurate diagnostic information for lung diseases. 

\section{Discussion}
\label{sec:discussion}
This manuscript has shown the first dynamic \textit{in vivo} dark-field CT lung imaging, achieved via the single-grid imaging technique. This technique simplifies dynamic imaging since it only requires a single sample exposure and a simple set-up, as opposed to other dark-field imaging techniques \cite{kitchen2020emphysema,velroyen2015grating,burkhardt2021vivo}, which require multiple sample exposures and/or a more complicated setup to extract the dark-field signal. Using single-grid imaging, we were able to measure the three-dimensional dark-field signal and thus the scattering strength of the lung tissues throughout a breath cycle via the linear diffusion coefficient, \(\mu_{d}\), which could potentially be used as a parameter to assess how healthy the lungs are, either globally or locally. In this section, we will discuss the feasibility of extracting quantitative measurements of alveolar dimension using this technique, optimal setup parameters, challenges in moving towards clinical application, and some future directions.  

\subsection{Quantitative dark-field signal} 
Since the strength of the dark-field signal originates from the alveoli of the lung, which are not fully resolved by the imaging system, this dark-field signal can provide complementary information about lung health. In the approach shown here, we see this complementarity and we can track the changes in the strength of the dark-field signal across each part of the lung over time. 

Future research could take the dark-field signals extracted from the lungs and quantify the average size of the alveoli, using calibration data obtained from PMMA microspheres of the same size as typical alveoli of mice, following the approach in \cite{how2023quantification}. In addition, the change in the surface-area-to-volume ratio of the alveoli could also be determined from the change in the retrieved \(\mu_{d}\) value, given the direct proportionality between these quantities \cite{paganin2023paraxial}. For example, if the \(\mu_{d}\) value increases by 20\%, then the surface-area-to-volume ratio of the alveoli has also increased by 20\% and vice versa. By tracking the change in the surface-area-to-volume ratio through the breath cycle, we could determine the expansion rate of the alveoli, which could provide a better assessment of lung function. Further analysis, including detailed comparisons with histology (see Supplementary Figures S1 - S3), could also be performed to better understand and verify how the dark-field signal is related to the size of the alveoli and hence lung health in various disease models. We hope that this quantitative time-resolved dark-field imaging technique can be developed into a new functional imaging modality and provide quantitative measurements for other lung diseases that also result in a change in the size and structure of the alveoli, such as fibrosis \cite{hellbach2017x} and COPD \cite{willer2021x}, which could be beneficial for biomedical research and eventually for clinical studies. 

\subsection{Optimal setup parameters to perform single-exposure dark-field imaging}\label{optimal_setup}
The key experimental parameters in this single-exposure x-ray dark-field imaging technique are the sample-to-detector propagation distance, the pixel size, and the period of the grid intensity pattern. Note that in this technique, the direct shadow of the grid on the detector is resolved, not the pattern that arises from the Talbot effect, and hence the expression for the Talbot distance \cite{paganin2006} is not applicable here. For the case of multiple refraction through the microstructures, if either a) the scattering cone angle $\theta$, or b) the sample microstructure size and material are known and the calibration curve (as shown in Fig.~7 of \cite{how2023quantification}) is available, then the optimal propagation distance can be calculated using \cite{how2023quantification}

\begin{equation}
        z_{opt} = \frac{p}{\pi \theta} = \frac{p}{\pi K \sqrt{T}}.
\end{equation}
Here, \(p\) represents the period of the grid intensity pattern, \(T\) represents the sample thickness which can be retrieved from the reconstructed attenuation CT slices, and the parameter \(K\) can be obtained from the calibration curve available for known microstructure size. However, while this could be determined for a class of samples (e.g.,~mouse lungs or rat lungs), an appropriate sample-to-detector propagation distance can also be determined experimentally. As a rule of thumb, the pixel size needs to be sufficiently small such that one period of the grid intensity pattern can be sampled by around 14 pixels (or at least 7 pixels), with some dependence on the point spread function of the camera. The blurring width from the sample microstructures needs to be a sufficiently large fraction of the point spread function that the blurring is visible. This can be achieved by changing the propagation distance to tune the sensitivity of the imaging system to the dark-field signal. The larger the propagation distance, the greater the blurring width and, thus, the stronger the dark-field signal. However, the propagation distance should not be too large to the extent that the grid intensity pattern is blurred-out altogether, which results in saturation of the dark-field signal. It is also worth noting that when moving to imaging larger animals, a longer propagation distance may be required if a larger pixel size is used, since a larger pixel is less sensitive to the dark-field blurring. 

\subsection{Challenges and future directions} 
Although the single-grid dark-field imaging technique only requires a single sample exposure, which is beneficial for dynamic imaging, the main challenge is that it is best-suited to mm-to-cm-sized samples since a relatively small pixel size is required to resolve blurring directly, as discussed in the previous section. Although this could be overcome by using a detector with both a large area and a small pixel size, such an approach may not be feasible due to the availability of such detectors. In addition, a relatively high radiation dose may be required to achieve sufficient counts in many small pixels.  

It is also worth noting that our algorithm is able to extract the differential phase signal in addition to the attenuation and dark-field signal. However, the differential phase signal extracted from the mouse was relatively weak compared to the other two signals, resulting in a noisy image. Since it is outside the scope of this paper, the differential phase images are not included. 

One possible direction for future investigation includes performing this technique using laboratory x-ray sources. Since the single-grid imaging technique only requires a certain degree of spatial coherence \cite{macindoe2016}, it could also be performed using a polychromatic x-ray source with a sufficiently small source size \cite{wen2010,macindoe2016} in a laboratory-based setup. Although this technique does not require a temporally coherent source, note that a polychromatic source can result in the inaccurate measurement of quantitative dark-field signal, due to beam hardening \cite{yashiro2015effect,pelzer2016beam,pandeshwar2020modeling} and visibility hardening \cite{de2023x} effects, which would result in an additional reduction to the visibility of the grid intensity pattern.  

To automate the process of tracking a particular volume of lung tissue through the breath, this technique could be combined with x-ray velocimetry (XV) analysis \cite{dubsky2012synchrotron}, which measures how much the lung tissues expand between each breath point \cite{stahr2016quantification,murrie2020real,werdiger2020quantification,reyne2024}. XV imaging could be performed on the phase-enhanced attenuation images, and hence automate the processing pipeline. 

In addition to \textit{in vivo} lung imaging, the 4D dark-field imaging technique described here could be useful for the imaging of other dynamic processes, such as the production of metal foams \cite{garcia2019using} or chemical reactions that involve the formation or decomposition of substances in an aqueous solution, which act as unresolved microstructures and provide a dark-field signal \cite{rieger2000study,albiter2003structural}. For these applications, the data acquisition protocol could be modified to perform multiple rotations (instead of just one rotation) to capture the non-periodic sample changes instead of the repeated sample changes seen in this study.     

The experimental approach taken here can also be used to extract a directional dark-field signal \cite{jensen2010a,jensen2010b}, which arises from microstructures that are elongated and aligned in a particular direction, in a dynamic process. By modelling the blurring function as a two-dimensional anisotropic Gaussian distribution \cite{jensen2010b}, Croughan \textit{et al}. \cite{croughan2022directional} successfully applied the single-grid technique to extract the directional dark-field signal, including both the angle at which the microstructures are oriented and the eccentricity of the blurring from the microstructures, from bundles of carbon fibres oriented in different directions with water slowly travelling from the bottom of the fibres to the top via capillary action (see Visualisation 1 from \cite{croughan2022directional}). This directional dark-field imaging technique can also be useful in imaging nanoparticle strings to monitor the delivery of drugs down the airways \cite{smith2024ultra}. 

Another challenge that we encountered in this experiment was the source-size blurring seen at larger sample-to-detector propagation distances. We observed that at a 3.5~m propagation distance, the grid intensity pattern in the horizontal direction was blurred significantly, with a resulting visibility of 0.08 (cf. 0.2 visibility in the vertical direction). This is due to the source being extended in the horizontal direction at the IMBL, resulting in a stronger source-size blurring, and thus it was more challenging to extract the dark-field signal in the horizontal direction. This could be overcome by performing a correction on the raw sample-and-grid images for these non-dark-field blurring effects prior to analysis, as demonstrated by Croughan \textit{et al.} \cite{croughan2024correcting}. This source-size blurring effect was not an issue in this application, since the dark-field signal from the alveoli was measured under the assumption of rotationally-symmetric x-ray scattering. However, this could be an issue for samples made up of microstructures that scatter anisotropically \cite{jensen2010a}. The inevitable motion-blurring effect from the heart was another issue we encountered in this experiment, but the effect was minimal since only a very small part of the lungs (i.e.,~the region surrounding the heart) is affected by this. With a high-brightness source, image capture could potentially also be synchronized to the heartbeat, as previously shown with propagation-based phase contrast imaging \cite{lovric2017tomographic}.   

\section{Conclusion} 
We have presented the first \textit{in vivo} 4D dark-field lung imaging of mice. This was achieved by synchronizing data acquisition with the breath cycle, and was possible because of the short acquisition times that can be used in single-grid dark-field imaging. This 4D imaging technique provides three-dimensional dark-field images with minimal motion blurring from the moving lungs throughout a breath cycle, a challenge with other dark-field imaging techniques that require multiple exposures to extract the dark-field signal. 

We have shown that the dark-field CT images provide information complementary to the attenuation CT images, for example, how the alveoli change in size throughout the breath cycle and how the size of the alveoli vary from one part of a lung to another, in the case of both healthy and diseased lungs. It was observed that the strength of the dark-field signal decreases during inhalation, suggesting that the alveoli are expanding, thus increasing in size; during exhalation, the strength of the dark-field signal increases, suggesting that the alveoli are contracting/relaxing, thus decreasing in size. The reconstructed dark-field signal, or more specifically, the linear diffusion coefficient value, \(\mu_{d}\), also indicates the surface-area-to-volume ratio of the alveoli within each voxel \cite{paganin2023paraxial}. We were also able to measure the dark-field signal from specific ROIs, which provides regional information for the lungs.  

One future direction includes quantitatively relating the time-resolved dark-field signal to the physical properties of the alveoli, such as the size or the rate of expansion. This could allow us to measure the expansion rate of the alveoli and thus regionally assess how healthy the lungs are. Furthermore, a future study could look at the variability in these measures across larger cohorts of animals. Although this technique is currently limited to the imaging of small-sized samples, we hope that this technique can equip biomedical research studies immediately and, long-term, potentially be developed into a low-dose, non-invasive and accurate functional lung imaging technique in the future.      

\section*{Supplementary Information}
Supplementary information can be found here: \\
\url{https://doi.org/10.26180/27886563.v1}


\bibliographystyle{IEEEtran} 
\bibliography{reference_list} 

\begin{thebibliography}{10}
\providecommand{\url}[1]{#1}
\csname url@samestyle\endcsname
\providecommand{\newblock}{\relax}
\providecommand{\bibinfo}[2]{#2}
\providecommand{\BIBentrySTDinterwordspacing}{\spaceskip=0pt\relax}
\providecommand{\BIBentryALTinterwordstretchfactor}{4}
\providecommand{\BIBentryALTinterwordspacing}{\spaceskip=\fontdimen2\font plus
\BIBentryALTinterwordstretchfactor\fontdimen3\font minus \fontdimen4\font\relax}
\providecommand{\BIBforeignlanguage}[2]{{%
\expandafter\ifx\csname l@#1\endcsname\relax
\typeout{** WARNING: IEEEtran.bst: No hyphenation pattern has been}%
\typeout{** loaded for the language `#1'. Using the pattern for}%
\typeout{** the default language instead.}%
\else
\language=\csname l@#1\endcsname
\fi
#2}}
\providecommand{\BIBdecl}{\relax}
\BIBdecl

\bibitem{paganin2006}
D.~M. Paganin, \emph{{Coherent X-Ray Optics}}.\hskip 1em plus 0.5em minus 0.4em\relax Oxford University Press, 2006.

\bibitem{rigon2007_abi_3}
L.~Rigon, F.~Arfelli, and R.-H. Menk, ``Generalized diffraction enhanced imaging to retrieve absorption, refraction and scattering effects,'' \emph{J. Phys. D Appl. Phys.}, vol.~40, no.~10, pp. 3077--3089, May 2007, \url{10.1088/0022-3727/40/10/011}.

\bibitem{pfeiffer2008}
F.~Pfeiffer \emph{et~al.}, ``Hard-{X}-ray dark-field imaging using a grating interferometer,'' \emph{Nat. Mater.}, vol.~7, no.~2, pp. 134--137, Feb. 2008, \url{10.1038/nmat2096}.

\bibitem{endrizzi2014ei}
M.~Endrizzi \emph{et~al.}, ``Hard {X}-ray dark-field imaging with incoherent sample illumination,'' \emph{Appl. Phys. Lett.}, vol. 104, no.~2, Jan. 2014, {A}rt. no. 024106, \url{10.1063/1.4861855}.

\bibitem{endrizzi2017x}
------, ``X-ray phase-contrast radiography and tomography with a multiaperture analyzer,'' \emph{Phys. Rev. Lett.}, vol. 118, no.~24, Jun. 2017, {A}rt. no. 243902, \url{10.1103/PhysRevLett.118.243902}.

\bibitem{berujon2012_sb}
\BIBentryALTinterwordspacing
S.~Berujon, H.~Wang, and K.~Sawhney, ``X-ray multimodal imaging using a random-phase object,'' \emph{Phys. Rev. A.}, vol.~86, no.~6, Dec. 2012. [Online]. Available: \url{{A}rt. no. 063813, \url{10.1103/PhysRevA.86.063813}}
\BIBentrySTDinterwordspacing

\bibitem{zanette2014}
I.~Zanette \emph{et~al.}, ``Speckle-based x-ray phase-contrast and dark-field imaging with a laboratory source,'' \emph{Phys. Rev. Lett.}, vol. 112, no.~25, Jun. 2014, {A}rt. no. 253903, \url{10.1103/PhysRevLett.112.253903}.

\bibitem{zdora2017x}
M.-C. Zdora \emph{et~al.}, ``X-ray phase-contrast imaging and metrology through unified modulated pattern analysis,'' \emph{Phys. Rev. Lett.}, vol. 118, no.~20, May 2017, {A}rt. no. 203903, \url{10.1103/PhysRevLett.118.203903}.

\bibitem{pavlov2020x}
K.~M. Pavlov, D.~M. Paganin, H.~T. Li, S.~Berujon, H.~Roug{\'e}-Labriet, and E.~Brun, ``X-ray multi-modal intrinsic-speckle-tracking,'' \emph{J. Opt.}, vol.~22, no.~12, Nov. 2020, {A}rt. no. 125604, \url{10.1088/2040-8986/abc313}.

\bibitem{alloo2022dark}
S.~J. Alloo \emph{et~al.}, ``Dark-field tomography of an attenuating object using intrinsic x-ray speckle tracking,'' \emph{J. Med. Imaging}, vol.~9, no.~3, May 2022, {A}rt. no. 031502, \url{10.1117/1.jmi.9.3.031502}.

\bibitem{beltran2023fast}
M.~A. Beltran, D.~M. Paganin, M.~K. Croughan, and K.~S. Morgan, ``Fast implicit diffusive dark-field retrieval for single-exposure, single-mask x-ray imaging,'' \emph{Optica}, vol.~10, no.~4, pp. 422--429, Apr. 2023, \url{10.1364/OPTICA.480489}.

\bibitem{gureyev2020}
T.~E. Gureyev \emph{et~al.}, ``Dark-field signal extraction in propagation-based phase-contrast imaging,'' \emph{Phys. Med. Biol.}, vol.~65, no.~21, Nov. 2020, {A}rt. no. 215029, \url{10.1088/1361-6560/abac9d}.

\bibitem{leatham2023x}
T.~A. Leatham, D.~M. Paganin, and K.~S. Morgan, ``X-ray dark-field and phase retrieval without optics, via the {F}okker--{P}lanck equation,'' \emph{IEEE Trans. Med. Imaging}, vol.~42, no.~6, pp. 1681--1695, Jan. 2023, \url{10.1109/TMI.2023.3234901}.

\bibitem{leatham2024x}
------, ``X-ray phase and dark-field computed tomography without optical elements,'' \emph{Opt. Express}, vol.~32, no.~3, pp. 4588--4602, Jan. 2024, \url{10.1364/OE.509604}.

\bibitem{ahlers2024}
J.~N. Ahlers, K.~M. Pavlov, M.~J. Kitchen, and K.~S. Morgan, ``X-ray dark-field via spectral propagation-based imaging,'' \emph{Optica}, vol.~11, no.~8, pp. 1182--1191, Aug. 2024, \url{10.1364/OPTICA.506742}.

\bibitem{wen2010}
H.~H. Wen, E.~E. Bennett, R.~Kopace, A.~F. Stein, and V.~Pai, ``Single-shot x-ray differential phase-contrast and diffraction imaging using two-dimensional transmission gratings,'' \emph{Opt. Lett.}, vol.~35, no.~12, pp. 1932--1934, Jun. 2010, \url{10.1364/OL.35.001932}.

\bibitem{morgan2013}
K.~S. Morgan \emph{et~al.}, ``A sensitive x-ray phase contrast technique for rapid imaging using a single phase grid analyzer,'' \emph{Opt. Lett.}, vol.~38, no.~22, pp. 4605--4608, Nov. 2013, \url{10.1364/OL.38.004605}.

\bibitem{croughan2022directional}
M.~K. Croughan, Y.~Y. How, A.~Pennings, and K.~S. Morgan, ``Directional dark-field retrieval with single-grid x-ray imaging,'' \emph{Opt. Express}, vol.~31, no.~7, pp. 11\,578--11\,597, Mar. 2023, \url{10.1364/OE.480031}.

\bibitem{kitchen2020emphysema}
M.~J. Kitchen \emph{et~al.}, ``Emphysema quantified: mapping regional airway dimensions using 2{D} phase contrast {X}-ray imaging,'' \emph{Biomed. Opt. Express}, vol.~11, no.~8, pp. 4176--4190, Aug. 2020, \url{10.1364/BOE.390587}.

\bibitem{bech2010_coeff}
M.~Bech, O.~Bunk, T.~Donath, R.~Feidenhans, C.~David, and F.~Pfeiffer, ``Quantitative x-ray dark-field computed tomography,'' \emph{Phys. Med. Biol.}, vol.~55, no.~18, pp. 5529--5539, Aug. 2010, \url{10.1088/0031-9155/55/18/017}.

\bibitem{lynch2011_coeff}
S.~K. Lynch \emph{et~al.}, ``Interpretation of dark-field contrast and particle-size selectivity in grating interferometers,'' \emph{Appl. Opt.}, vol.~50, no.~22, pp. 4310--4319, Jul. 2011, \url{10.1364/AO.50.004310}.

\bibitem{gkoumas2016}
S.~Gkoumas, P.~Villanueva-Perez, Z.~Wang, L.~Romano, M.~Abis, and M.~Stampanoni, ``A generalized quantitative interpretation of dark-field contrast for highly concentrated microsphere suspensions,'' \emph{Sci. Rep.}, vol.~6, Oct. 2016, {A}rt. no. 35259, \url{10.1038/srep35259}.

\bibitem{prade2016}
F.~Prade, A.~Yaroshenko, J.~Herzen, and F.~Pfeiffer, ``Short-range order in mesoscale systems probed by {X}-ray grating interferometry,'' \emph{Europhys. Lett.}, vol. 112, no.~6, Jan. 2016, {A}rt. no. 68002, \url{10.1209/0295-5075/112/68002}.

\bibitem{harti2017}
R.~P. Harti, M.~Strobl, B.~Betz, K.~Jefimovs, M.~Kagias, and C.~Grünzweig, ``Sub-pixel correlation length neutron imaging: Spatially resolved scattering information of microstructures on a macroscopic scale,'' \emph{Sci. Rep.}, vol.~7, no.~1, Mar. 2017, {A}rt. no. 44588, \url{10.1038/srep44588}.

\bibitem{modregger2017}
P.~Modregger \emph{et~al.}, ``Interpretation and utility of the moments of small-angle x-ray scattering distributions,'' \emph{Phys. Rev. Lett.}, vol. 118, no.~26, Jun. 2017, {A}rt. no. 265501, \url{10.1103/physrevlett.118.265501}.

\bibitem{how2022quantifying}
Y.~Y. How and K.~S. Morgan, ``Quantifying the x-ray dark-field signal in single-grid imaging,'' \emph{Opt. Express}, vol.~30, no.~7, pp. 10\,899--10\,918, Mar. 2022, \url{10.1364/OE.451834}.

\bibitem{how2023quantification}
Y.~Y. How, D.~M. Paganin, and K.~S. Morgan, ``On the quantification of sample microstructure using single-exposure x-ray dark-field imaging via a single-grid setup,'' \emph{Sci. Rep.}, vol.~13, no.~1, Jul. 2023, {A}rt. no. 11001, \url{10.1038/s41598-023-37334-3}.

\bibitem{yashiro2010origin}
W.~Yashiro, Y.~Terui, K.~Kawabata, and A.~Momose, ``On the origin of visibility contrast in x-ray {T}albot interferometry,'' \emph{Opt. Express}, vol.~18, no.~16, pp. 16\,890--16\,901, Aug. 2010, \url{10.1364/OE.18.016890}.

\bibitem{miller2013phase}
E.~A. Miller, T.~A. White, B.~S. McDonald, and A.~Seifert, ``Phase contrast x-ray imaging signatures for security applications,'' \emph{IEEE Trans. Nucl. Sci.}, vol.~60, no.~1, pp. 416--422, Jan. 2013, \url{10.1109/TNS.2012.2227803}.

\bibitem{partridge2022enhanced}
T.~Partridge \emph{et~al.}, ``Enhanced detection of threat materials by dark-field x-ray imaging combined with deep neural networks,'' \emph{Nat. Commun.}, vol.~13, no.~1, Sep. 2022, {A}rt. no. 4651, \url{10.1038/s41467-022-32402-0}.

\bibitem{valsecchi2020}
J.~Valsecchi \emph{et~al.}, ``Characterization of oriented microstructures through anisotropic small-angle scattering by 2{D} neutron dark-field imaging,'' \emph{Commun. Phys.}, vol.~3, no.~1, Feb. 2020, {A}rt. no. 42, \url{10.1038/s42005-020-0308-4}.

\bibitem{scherer2015non}
K.~Scherer \emph{et~al.}, ``Non-invasive differentiation of kidney stone types using {X}-ray dark-field radiography,'' \emph{Sci. Rep.}, vol.~5, no.~1, Apr. 2015, {A}rt. no. 9527, \url{10.1038/srep09527}.

\bibitem{niemann2021classification}
T.~Niemann, I.~Jerjen, L.~Hefermehl, Z.~Wang, R.~A. Kubik-Huch, and M.~Stampanoni, ``The classification of renal stones by gratings-based dark-field radiography,'' \emph{Cent. European J. Urol.}, vol.~74, no.~3, Sep. 2021, {A}rt. no. 453, \url{10.5173/ceju.2021.3.0334}.

\bibitem{michel2013dark}
T.~Michel \emph{et~al.}, ``On a dark-field signal generated by micrometer-sized calcifications in phase-contrast mammography,'' \emph{Phys. Med. Biol.}, vol.~58, no.~8, pp. 2713--2732, Apr. 2013, \url{10.1088/0031-9155/58/8/2713}.

\bibitem{wang2014non}
Z.~Wang \emph{et~al.}, ``Non-invasive classification of microcalcifications with phase-contrast {X}-ray mammography,'' \emph{Nat. Commun.}, vol.~5, no.~1, May 2014, {A}rt. no. 3797, \url{10.1038/ncomms4797}.

\bibitem{forte2020can}
S.~Forte \emph{et~al.}, ``Can grating interferometry-based mammography discriminate benign from malignant microcalcifications in fresh biopsy samples?'' \emph{Eur. J. Radiol.}, vol. 129, Aug. 2020, {A}rt. no. 109077, \url{10.1016/j.ejrad.2020.109077}.

\bibitem{aminzadeh2022}
A.~Aminzadeh \emph{et~al.}, ``Imaging breast microcalcifications using dark-field signal in propagation-based phase-contrast tomography,'' \emph{IEEE Trans. Med. Imaging}, vol.~41, no.~11, pp. 2980--2990, May 2022, \url{10.1109/TMI.2022.3175924}.

\bibitem{gassert2023dark}
F.~T. Gassert \emph{et~al.}, ``Dark-field x-ray imaging for the assessment of osteoporosis in human lumbar spine specimens,'' \emph{Front. Psychol.}, vol.~14, Jul. 2023, {A}rt. no. 1217007, \url{10.3389/fphys.2023.1217007}.

\bibitem{schaff2024feasibility}
F.~Schaff \emph{et~al.}, ``Feasibility of dark-field radiography to enhance detection of nondisplaced fractures,'' \emph{Radiology}, vol. 311, no.~2, May 2024, {A}rt. no. e231921, \url{10.1148/radiol.231921}.

\bibitem{schleede2012emphysema}
S.~Schleede, F.~G. Meinel, M.~Bech, J.~Herzen, K.~Achterhold, G.~Potdevin, A.~Malecki, S.~Adam-Neumair, S.~F. Thieme, F.~Bamberg \emph{et~al.}, ``Emphysema diagnosis using x-ray dark-field imaging at a laser-driven compact synchrotron light source,'' \emph{Proc. Natl. Acad. Sci.}, vol. 109, no.~44, pp. 17\,880--17\,885, Oct. 2012, \url{10.1073/pnas.1206684109}.

\bibitem{yaroshenko2013pulmonary}
A.~Yaroshenko \emph{et~al.}, ``Pulmonary emphysema diagnosis with a preclinical small-animal x-ray dark-field scatter-contrast scanner,'' \emph{Radiology}, vol. 269, no.~2, pp. 427--433, Nov. 2013, \url{10.1148/radiol.13122413}.

\bibitem{scherer2017x}
K.~Scherer \emph{et~al.}, ``X-ray dark-field radiography -- \textit{in-vivo} diagnosis of lung cancer in mice,'' \emph{Sci. Rep.}, vol.~7, no.~1, Mar. 2017, {A}rt. no. 402, \url{10.1038/s41598-017-00489-x}.

\bibitem{hellbach2015vivo}
K.~Hellbach \emph{et~al.}, ``In vivo dark-field radiography for early diagnosis and staging of pulmonary emphysema,'' \emph{Invest. Radiol.}, vol.~50, no.~7, pp. 430--435, Jul. 2015, \url{10.1097/rli.0000000000000147}.

\bibitem{hellbach2017x}
------, ``X-ray dark-field radiography facilitates the diagnosis of pulmonary fibrosis in a mouse model,'' \emph{Sci. Rep.}, vol.~7, no.~1, Mar. 2017, {A}rt. no. 340, \url{10.1038/s41598-017-00475-3}.

\bibitem{gassert2021x}
F.~T. Gassert \emph{et~al.}, ``X-ray dark-field chest imaging: qualitative and quantitative results in healthy humans,'' \emph{Radiology}, vol. 301, pp. 389--395, Nov. 2021, \url{10.1148/radiol.2021210963}.

\bibitem{frank2022dark}
M.~Frank \emph{et~al.}, ``Dark-field chest x-ray imaging for the assessment of {COVID}-19-pneumonia,'' \emph{Commun. Med.}, vol.~2, no.~1, Nov. 2022, {A}rt. no. 147, \url{10.1038/s43856-022-00215-3}.

\bibitem{willer2021x}
K.~Willer \emph{et~al.}, ``X-ray dark-field chest imaging for detection and quantification of emphysema in patients with chronic obstructive pulmonary disease: a diagnostic accuracy study,'' \emph{Lancet Digit Health}, vol.~3, no.~11, pp. e733--e744, Nov. 2021, \url{10.1016/s2589-7500(21)00146-1}.

\bibitem{urban2023dark}
T.~Urban \emph{et~al.}, ``Dark-field chest radiography outperforms conventional chest radiography for the diagnosis and staging of pulmonary emphysema,'' \emph{Invest. Radiol.}, vol.~58, no.~11, pp. 775--781, Nov. 2023, \url{10.1097/RLI.0000000000000989}.

\bibitem{viermetz2022dark}
M.~Viermetz \emph{et~al.}, ``Dark-field computed tomography reaches the human scale,'' \emph{Proc. Natl. Acad. Sci.}, vol. 119, no.~8, Feb. 2022, {A}rt. no. e2118799119, \url{10.1073/pnas.2118799119}.

\bibitem{meinel2013diagnosing}
F.~G. Meinel \emph{et~al.}, ``Diagnosing and mapping pulmonary emphysema on {X}-ray projection images: incremental value of grating-based {X}-ray dark-field imaging,'' \emph{PloS ONE}, vol.~8, no.~3, Mar. 2013, {A}rt. no. e59526, \url{10.1371/journal.pone.0059526}.

\bibitem{gromann2017vivo}
L.~B. Gromann \emph{et~al.}, ``In-vivo x-ray dark-field chest radiography of a pig,'' \emph{Sci. Rep.}, vol.~7, no.~1, Jul. 2017, {A}rt. no. 4807, \url{10.1038/s41598-017-05101-w}.

\bibitem{gradl2018dynamic}
R.~Gradl \emph{et~al.}, ``Dynamic in vivo chest x-ray dark-field imaging in mice,'' \emph{IEEE Trans. Med. Imaging}, vol.~38, no.~2, pp. 649--656, Sep. 2018, \url{10.1109/tmi.2018.2868999}.

\bibitem{morgan2011_grid}
K.~S. Morgan, D.~M. Paganin, and K.~K. Siu, ``Quantitative single-exposure x-ray phase contrast imaging using a single attenuation grid,'' \emph{Opt. {E}xpress}, vol.~19, no.~20, pp. 19\,781--19\,789, Sep. 2011, \url{10.1364/OE.19.019781}.

\bibitem{mall2004increased}
M.~Mall, B.~R. Grubb, J.~R. Harkema, W.~K. O'Neal, and R.~C. Boucher, ``Increased airway epithelial na+ absorption produces cystic fibrosis-like lung disease in mice,'' \emph{Nat. Med.}, vol.~10, no.~5, pp. 487--493, May 2004, \url{10.1038/nm1028}.

\bibitem{mall2008development}
M.~A. Mall \emph{et~al.}, ``Development of chronic bronchitis and emphysema in $\beta$-epithelial {N}a\textsuperscript{+} channel--overexpressing mice,'' \emph{Am. J. Respir. Crit. Care Med.}, vol. 177, no.~7, pp. 730--742, Apr. 2008, \url{10.1164/rccm.200708-1233OC}.

\bibitem{zhou2017airway}
Z.~Zhou-Suckow, J.~Duerr, M.~Hagner, R.~Agrawal, and M.~A. Mall, ``Airway mucus, inflammation and remodeling: emerging links in the pathogenesis of chronic lung diseases,'' \emph{Cell Tissue Res.}, vol. 367, pp. 537--550, Mar. 2017, \url{10.1007/s00441-016-2562-z}.

\bibitem{wielputz2011vivo}
M.~O. Wielp{\"u}tz \emph{et~al.}, ``In vivo monitoring of cystic fibrosis-like lung disease in mice by volumetric computed tomography,'' \emph{Eur. Respir. J.}, vol.~38, no.~5, pp. 1060--1070, Nov. 2011, \url{10.1183/09031936.00149810}.

\bibitem{mccarron2021animal}
A.~McCarron, D.~Parsons, and M.~Donnelley, ``Animal and cell culture models for cystic fibrosis: which model is right for your application?'' \emph{Am. J. Pathol.}, vol. 191, no.~2, pp. 228--242, Feb. 2021, \url{10.1016/j.ajpath.2020.10.017}.

\bibitem{zhu2022muct}
L.~Zhu \emph{et~al.}, ``\textmu{CT} to quantify muco-obstructive lung disease and effects of neutrophil elastase knockout in mice,'' \emph{Am. J. Physiol. Lung Cell. Mol. Physiol.}, vol. 322, no.~3, pp. L401--L411, Mar. 2022, \url{10.1152/ajplung.00341.2021}.

\bibitem{reyne2024}
N.~Reyne \emph{et~al.}, ``Functional lung imaging identifies peripheral ventilation changes in mice with muco-obstructive lung disease,'' \emph{bioRxiv}, Jul. 2024, \url{10.1101/2024.06.27.600946}.

\bibitem{weeden2023early}
C.~E. Weeden \emph{et~al.}, ``Early immune pressure initiated by tissue-resident memory {T} cells sculpts tumor evolution in non-small cell lung cancer,'' \emph{Cancer Cell}, vol.~41, no.~5, pp. 837--852, May 2023, \url{10.1016/j.ccell.2023.03.019}.

\bibitem{morgan2020methods}
K.~S. Morgan \emph{et~al.}, ``Methods for dynamic synchrotron {X}-ray respiratory imaging in live animals,'' \emph{J. Synchrotron Radiat.}, vol.~27, no.~1, pp. 164--175, Jan. 2020, \url{10.1107/S1600577519014863}.

\bibitem{van2015astra}
W.~Van~Aarle \emph{et~al.}, ``The {ASTRA} toolbox: A platform for advanced algorithm development in electron tomography,'' \emph{Ultramicroscopy}, vol. 157, pp. 35--47, Oct. 2015, \url{10.1016/j.ultramic.2015.05.002}.

\bibitem{van2016fast}
------, ``Fast and flexible x-ray tomography using the {ASTRA} toolbox,'' \emph{Opt. Express}, vol.~24, no.~22, pp. 25\,129--25\,147, Oct. 2016, \url{10.1364/OE.24.025129}.

\bibitem{velroyen2015grating}
A.~Velroyen \emph{et~al.}, ``Grating-based x-ray dark-field computed tomography of living mice,'' \emph{EBioMedicine}, vol.~2, no.~10, pp. 1500--1506, Oct. 2015, \url{10.1016/j.ebiom.2015.08.014}.

\bibitem{burkhardt2021vivo}
R.~Burkhardt \emph{et~al.}, ``In-vivo x-ray dark-field computed tomography for the detection of radiation-induced lung damage in mice,'' \emph{Phys. Imaging Radiat. Oncol.}, vol.~20, pp. 11--16, Oct. 2021, \url{10.1016/j.phro.2021.09.003}.

\bibitem{paganin2023paraxial}
D.~M. Paganin, D.~Pelliccia, and K.~S. Morgan, ``Paraxial diffusion-field retrieval,'' \emph{Phys. Rev. A.}, vol. 108, no.~1, Jul. 2023, {A}rt. no. 013517, \url{10.1103/PhysRevA.108.013517}.

\bibitem{macindoe2016}
D.~Macindoe, M.~J. Kitchen, S.~C. Irvine, A.~Fouras, and K.~S. Morgan, ``Requirements for dynamical differential phase contrast x-ray imaging with a laboratory source,'' \emph{Phys. Med. Biol.}, vol.~61, no.~24, pp. 8720--8735, Nov. 2016, \url{10.1088/1361-6560/61/24/8720}.

\bibitem{yashiro2015effect}
W.~Yashiro, P.~Vagovi{\v{c}}, and A.~Momose, ``Effect of beam hardening on a visibility-contrast image obtained by {X}-ray grating interferometry,'' \emph{Opt. Express}, vol.~23, no.~18, pp. 23\,462--23\,471, Sep. 2015, \url{10.1364/OE.23.023462}.

\bibitem{pelzer2016beam}
G.~Pelzer \emph{et~al.}, ``A beam hardening and dispersion correction for x-ray dark-field radiography,'' \emph{Med. Phys.}, vol.~43, no.~6, pp. 2774--2779, Jun. 2016, \url{10.1118/1.4948671}.

\bibitem{pandeshwar2020modeling}
A.~Pandeshwar, M.~Kagias, Z.~Wang, and M.~Stampanoni, ``Modeling of beam hardening effects in a dual-phase {X}-ray grating interferometer for quantitative dark-field imaging,'' \emph{Opt. Express}, vol.~28, no.~13, pp. 19\,187--19\,204, Jun. 2020, \url{10.1364/OE.395237}.

\bibitem{de2023x}
F.~De~Marco \emph{et~al.}, ``X-ray dark-field signal reduction due to hardening of the visibility spectrum,'' \emph{IEEE Trans. Med. Imaging}, vol.~43, no.~4, pp. 1422--1433, Nov. 2024, \url{10.1109/TMI.2023.3337994}.

\bibitem{dubsky2012synchrotron}
S.~Dubsky, S.~B. Hooper, K.~K.~W. Siu, and A.~Fouras, ``Synchrotron-based dynamic computed tomography of tissue motion for regional lung function measurement,'' \emph{J. R. Soc. Interface}, vol.~9, no.~74, pp. 2213--2224, Sep. 2012, \url{10.1098/rsif.2012.0116}.

\bibitem{stahr2016quantification}
C.~S. Stahr \emph{et~al.}, ``Quantification of heterogeneity in lung disease with image-based pulmonary function testing,'' \emph{Sci. Rep.}, vol.~6, no.~1, Jul. 2016, {A}rt. no. 29438, \url{10.1038/srep29438}.

\bibitem{murrie2020real}
R.~P. Murrie \emph{et~al.}, ``Real-time in vivo imaging of regional lung function in a mouse model of cystic fibrosis on a laboratory {X}-ray source,'' \emph{Sci. Rep.}, vol.~10, no.~1, Jan. 2020, {A}rt. no. 447, \url{10.1038/s41598-019-57376-w}.

\bibitem{werdiger2020quantification}
F.~Werdiger \emph{et~al.}, ``Quantification of muco-obstructive lung disease variability in mice via laboratory x-ray velocimetry,'' \emph{Sci. Rep.}, vol.~10, no.~1, Jul. 2020, {A}rt. no. 10859, \url{10.1038/s41598-020-67633-y}.

\bibitem{garcia2019using}
F.~Garc{\'\i}a-Moreno \emph{et~al.}, ``Using {X}-ray tomoscopy to explore the dynamics of foaming metal,'' \emph{Nat. Commun.}, vol.~10, no.~1, Aug. 2019, {A}rt. no. 3762, \url{10.1038/s41467-019-11521-1}.

\bibitem{rieger2000study}
J.~Rieger, J.~Thieme, and C.~Schmidt, ``Study of precipitation reactions by {X}-ray microscopy: {C}a{CO}$_3$ precipitation and the effect of polycarboxylates,'' \emph{Langmuir}, vol.~16, no.~22, pp. 8300--8305, Oct. 2000, \url{10.1021/la0004193}.

\bibitem{albiter2003structural}
A.~Albiter, A.~Contreras, E.~Bedolla, and R.~Perez, ``Structural and chemical characterization of precipitates in {Al}-2024/{TiC} composites,'' \emph{Compos. - A: Appl. Sci. Manuf.}, vol.~34, no.~1, pp. 17--24, Jan. 2003, \url{10.1016/S1359-835X(02)00259-2}.

\bibitem{jensen2010a}
T.~H. Jensen \emph{et~al.}, ``Directional x-ray dark-field imaging,'' \emph{Phys. Med. Biol.}, vol.~55, no.~12, pp. 3317--3323, May 2010, \url{10.1088/0031-9155/55/12/004}.

\bibitem{jensen2010b}
------, ``Directional x-ray dark-field imaging of strongly ordered systems,'' \emph{Phys. Rev. B.}, vol.~82, Dec. 2010, {A}rt. no. 214103, \url{10.1103/PhysRevB.82.214103}.

\bibitem{smith2024ultra}
R.~Smith \emph{et~al.}, ``Ultra-fast in vivo directional dark-field x-ray imaging for visualising magnetic control of particles for airway gene delivery,'' \emph{Phys. Med. Biol.}, vol.~69, no.~10, May 2024, {A}rt. no. 105025, \url{10.1088/1361-6560/ad40f5}.

\bibitem{croughan2024correcting}
M.~K. Croughan \emph{et~al.}, ``Correcting directional dark field x-ray imaging artefacts using position dependent image deblurring and attenuation removal,'' \emph{Sci. Rep.}, vol.~14, no.~1, Aug. 2024, {A}rt. no. 17807, \url{10.1038/s41598-024-68659-2}.

\bibitem{lovric2017tomographic}
G.~Lovric \emph{et~al.}, ``Tomographic in vivo microscopy for the study of lung physiology at the alveolar level,'' \emph{Sci. Rep.}, vol.~7, no.~1, Oct. 2017, {A}rt. no. 12545, \url{10.1038/s41598-017-12886-3}.

\end{thebibliography}

\end{document}